\documentstyle[multicol,aps,epsfig,psfig]{revtex}

\begin{document}

\draft

\title{On the form of growing strings}

\author{D. Marenduzzo$^{1}$, T.~X. Hoang$^{2}$, F. Seno$^{3,4}$, 
M. Vendruscolo$^5$, A. Maritan$^{3,4}$ }


\address{
$^1$ Mathematics Institute, University of Warwick, Coventry CV4 7AL,
England \\
$^2$Institute of Physics and Electronics, Vietnamese Academy
of Science and Technology, 10 Dao Tan, Hanoi, Viet Nam \\
$^3$INFM and Dipartimento di Fisica--Universit\`a di Padova, Via
Marzolo 8, 35131 Padova, Italy \\
$^4$ Sezione INFN, Universita' di Padova, I-35131 Padova, Italy \\
$^5$Department of Chemistry, University of Cambridge, Lensfield Road,
Cambridge CB2 1EW, England}

\maketitle

\begin{abstract}

Patterns and forms adopted by Nature, such as the shape of living cells, 
the geometry of shells and the branched structure of plants, are often 
the result of simple dynamical paradigms\cite{thompson}.
Here we show that a growing self-interacting string attached to a 
tracking origin, modeled to resemble nascent polypeptides 
{\em in vivo}, develops helical structures 
which are more pronounced at the growing end. We also show that 
the dynamic growth ensemble shares several features of an 
equilibrium ensemble in which the growing end of the polymer 
is under an effective stretching force.
A statistical analysis of native states of proteins shows that
the signature of this non-equilibrium phenomenon has been fixed by
evolution at the C-terminus, the growing end of a nascent protein. 
These findings suggest that a generic non-equilibrium growth process
might have provided an additional evolutionary advantage for nascent 
proteins by favoring the preferential selection of helical structures.
\end{abstract}

\begin{multicols}{2}


Non-equilibrium and pattern formation studies have a profound influence 
on {\em in vivo} biology \cite{thompson}. 
The dynamical evolution of the branched actin cytoskeletal network
which pushes a cell forward ~\cite{cytoskeleton},  
the spatial self-organisation of genome during 
DNA replication and transcription \cite{alberts}, 
and the synthesis of a nascent protein within a ribosome 
\cite{berisio,cell2} are just a few examples in which 
non-equilibrium physics becomes relevant to cell biology.
In order to make contact with these phenomena occurring within
the cell, {\em in vitro} 
experiments on polymer dynamics employing single molecule techniques
are nowadays routinely made. These are 
amenable to a detailed analysis and to comparison with theory. 
Examples include DNA packaging in the $\Phi$29 bacteriophage \cite{phi29}, 
DNA or polymer translocation through a membrane pore \cite{translocation1},
the rheology of a single molecule of DNA under shear
or elongational flow \cite{chu} and the elasticity of 
F-actin networks \cite{cytoskeletonrheology}. 
These experiments draw on the ability to micro-manipulate
biopolymers with high accuracy, and even at the single molecule level
\cite{single}, employing atomic force microscopes, soft micro-needles,
laser tweezers and  other micro or nanoscopic devices.

Despite their importance, studies in 
polymer dynamics out of or far from equilibrium 
pose serious theoretical challenges and 
have not been systematically investigated so far.
Here we report evidence for the existence of a 
non-equilibrium pattern selection favoring the
formation of helices during the growth of a 
self-interacting polymer from a tracking origin, 
mimicking the situation occurring when nascent 
proteins are produced\cite{berisio,cell2}. 
We show by numerical simulations that the 
dynamic ensemble we consider shares many features 
of an equilibrium ensemble in which the
newly grown polymer segment is under a stretching force,
and we discuss where the "dynamic" force comes from in our simulations.
Finally, by an analysis of protein structures from the 
Protein-Data-Bank, we observe that the dynamic preference 
for helices we pinpoint has been fixed by evolution in the folded
structure of proteins and we speculate why it might be so.


We modelled the growth of a flexible self-interacting polymer
attached to a moving, or tracking, origin by means of 
molecular dynamics \cite{hoangjcp}.
We work in the reference frame of the tracking origin.
The polymer is modelled by a chain of $N$ beads, at positions  
$\left\{\vec r_i\right\}_{i=1,\ldots,N}$ in the 
three-dimensional space. The growth process starts from a 
conformation which contains a single bead placed at the origin of the 
reference frame.
After a repeated constant time interval, a new bead
is added at the position of the growing end and the rest of the
chain is translated upward (always in the positive $z$-direction) 
by one link.
The polymeric beads are tethered by a harmonic potential
between any two consecutive residues, given by
\begin{equation}
V(r) = k (r-b)^2 \;,
\end{equation}
with $r$ their mutual distance, $b$ the equilibrium length of a link and $k$
the spring constant. Non-consecutive residues interact via a 6-12 Lennard-Jones
(LJ) potential mimicking a poor solvent,
\begin{equation}\label{LJ}
V_{LJ}(r)=4\epsilon\left(
\left(\frac{\sigma}{r}\right)^{12}-\left(\frac{\sigma}{r}\right)^{6} \right),
\end{equation}
where $\sigma$ and $\epsilon$, determining the range and strength of the LJ
potential, also define the length and energy scales in our model.  The unit of
time is $\tau \equiv \sqrt{m\sigma^2/\epsilon}$ where $m$ is the mass,
identical for every residue.  We have chosen $b=0.78\sigma$, so that the ratio
$\sigma/b$ is close to that of a polypeptide chain, and
$k=2500\epsilon/\sigma^2$.
Coupling of the chain with the surrounding solution is given by a damping
term and the Langevin noise that compensates the drag force in order to
maintain constant temperature, $T$. The equations of motion for each bead are
\begin{equation}
m \ddot{\vec r_i} = -\gamma \dot{\vec r_i} + {\vec F_c} + {\vec \Gamma},
\end{equation}
where ${\vec F_c}$ is the net force due to molecular potentials 
(1) and (2), 
$\gamma$ is
the friction constant and ${\vec \Gamma}$ is a Gaussian noise with dispersion
$\sqrt{2\gamma k_B T}$. We have chosen $\gamma = 10 m/\tau$ which
is above the overdamping limit \cite{folding_times1}. 
The equations of motion are
integrated by means of the fifth-order  
Gear predictor-corrector algorithm \cite{gear}
with a time step of 0.005$\tau$.
The simulations are carried out at near zero temperature, $T=0.02\epsilon/k_B$,
in order to provide a highly stabilizing condition. The growth rate, $r_{growth}$,
is measured as number of beads grown per unit time.


The steady state regime of the configurations resulting from
the growth simulations is characterised by a dynamic coexistence
of polymer segments - or blobs - which are associated with a various degree 
of equilibration. Due to the
attractive self-interactions, mimicking van der Waals forces
in polypeptides, the patterns near the growing end depend
crucially on the ratio between the growth and the relaxation rates.
Representative configurations obtained by molecular dynamics simulations of the
growth process are shown in Fig. 1. If the growth rate, $r_{growth}$, is
small compared to the relaxation rate, $\tau_{\rm relax}^{-1}$, 
($\tau_{\rm relax}$ being the
characteristic time needed for the system to equilibrate) the string has 
the time to equilibrate and a rather compact and structureless 
conformation is obtained (Fig. 1d). 
The most interesting situation occurs when the two rates are comparable, 
$r_{growth} \tau_{relax}\sim 1$ (Fig. 1a--c). 
Near the growing end the polymer self-organises into a 
helix or a zig-zag structure, with its axis parallel to 
the growth direction. As string segments age and get farther
away from the growing end, they can organise themselves into
alternative shapes corresponding to different free energy minima. 
For large growing rates, $r_{\rm growth}\tau_{\rm relax}\gg 1$
the polymer is stretched as if under the action of a pulling force.
We consider two cases: (i) the growing end is kept fixed 
at the origin of the reference frame
(this more closely mimics the situation of a nascent peptide inside the 
ribosomal exit tunnel), and (ii) the growing end is free. 
While in both cases the results are qualitatively similar, in the former case
(shown in Fig. 1) the helix is longer and it is
formed over a wider range of $r_{growth}$. 
In both cases, the ranges of stability of the helix and the zig-zag
structure are enhanced as T decreases, i.e. they can form at slower
and wider ranges of the growth rates. This is mainly due to the
enhancement of the free energy barrier to be crossed to enter the globule phase
which effectively raises $\tau_{relax}$.

The phenomenon of pattern selection that we observe is mainly caused by 
the lack of time for the growing portion
of the chain to fold into a globule, so that only local contacts
along the string can be established. 
However, \textit{a priori}, other different periodic or
less regular structures with local contacts are possible during the
growth and thus the selection of a particular helix is rather a
surprise. The pitch to radius ratio of the helix depends on the
details of the self-interaction potential used for the string (in our
case the ratio between $\sigma$ and $b$). 
In Fig. 1c, instead, the aging end is becoming less ordered and more
compact than the helix at the growing end. 
It should be noted that the relaxation time of a polymer is
expected to have a power law dependence on the number of residues
$N$, $\tau_{\rm relax} \sim \tau_0 N^\alpha$, where $\alpha = 2$ if the
dynamics is Brownian and $\alpha \approx 1.7$ if long range
hydrodynamic interactions dominate the dynamics \cite{edwards}. Thus,
given any growth rate, if we wait long enough, the number of grown
beads increases and $r_{\rm growth} \tau_{\rm relax}$ will  eventually become
comparable to 1.


It is instructive to compare the conformations observed 
during the growth of a polymer (Fig. 1) with those 
resulting from simulated pulling experiments (Fig. 2).
We consider the same inter-bead potential (1) and (2) used for the
growth simulations. We considered two 
different ensembles in which a simulated pulling experiment
can be performed: (a) a 'fixed force' mode in which the string is
subject to a constant force and (b) a 'fixed stretch' mode in
which one end of the string is fixed and the other one is moved.
with a very low velocity so that the string is always found in
a near equilibrium condition.
In case (a) the controlled parameter is the magnitude of the pulling
force, $f$,  while in case (b) it is the stretching velocity $v$. 
Fig. 2 shows the set-up and the sketched `phase diagrams'
with representative ground state structures 
obtained with simulations of string pulling in modes (a) and (b). 
These `phase diagrams' could be tested with single molecule
stretching experiments.
This suggests that the dynamic growth ensemble
shares the qualitative features of a fixed force ensemble
near the newly grown portion of the polymer \cite{blob}. 


In our growth simulations, this effective ``stretching'' at 
the growing end 
is due both to the drag exerted by the more globular and
equilibrated ``older'' sections of the string onto the newly grown
ones, and to the inability of the newly grown segment to equilibrate
before a new segment of the string is grown. 
The growth mechanism that we consider resembles the one occurring 
{\em in vivo} when a nascent protein (the "self-attracting string") 
is synthesised in the ribosome (the "origin",
which is tracking along the mRNA).
The results that we presented show that non-equilibrium effects 
lead to specific pattern selection already in a generic 
self-interacting polymer. 
{This kinetic effect does not compete with other
well-known thermodynamic factors which
stabilizes $\alpha$-helices in proteins, such as
hydrogen bonds\cite{Pauling} and steric interactions\cite{Rama}.}
Furthermore, stereochemical considerations suggest that the
nascent protein can exit from the tunnel in an $\alpha$ helical 
conformation \cite{Baldwin}. The dynamic selection of helices 
may couple to the chemistry to yield a robust mechanism 
for helices formation during the very early stages of protein
folding. Still, what may the dynamic "stretching" come from during 
{\it in vivo} protein translation ? 
The ribosomal translation rate is about 45 or so nucleotides 
per second \cite{Baldwin}. Viscous drags would then 
be $\ll$ 1 pN for objects of the size of proteins (the
viscosity of cytosol for proteins is $\sim 1-10$ 
cP~\cite{cytosol}) so they are unlikely to be responsible 
for any co-translational stretching.
On the other hand, protein folding times are
typically smaller than even the total growth time.
In particular, $\beta$ sheets \cite{note} fold 
on the millisecond timescale \cite{folding_times1}, 
while the inverse growth rate is $\sim$ $0.1$ s (see above).
In narrow channel, however, such as the ribosomal exit tunnel 
(as tight as 2 nm in diameter, and 10 nm long, see \cite{berisio}), 
equilibration times increase significantly \cite{prl},
and $\beta$ formation may become sterically hindered, 
so that the ratio of these two numbers may match 
that seen in our simulations when helices are formed. 
Furthermore, once a helix is formed 
during the very early stages of a newly grown segment,
disrupting it will involve going over a free energy barrier $\Delta F$
so that the formation of more equilibrated structures will require
much longer (exponential in $\Delta F/T$)
than in an {\it in vitro} experiment, where the folding has
not been {\it a priori} ``biased'' towards any local minimum.


Finally, we have analysed the protein structures in the Protein Data Bank
(PDB) \cite{PDB} in order to verify whether the non-equilibrium 
phenomenon predicted here has left a signature even in the native state
of proteins, which is usually assumed to be determined by equilibrium
considerations alone.
We considered protein structures determined by X-ray diffraction 
and with less than {30\%} sequence identity, in order to remove 
homologous structures \cite{sander}. 
Furthermore, we removed proteins that are significantly disordered at 
their termini, i.e. when more than five residues from either terminus are 
unstructured. The propensities of forming $\alpha$-helix and
$\beta$-strand as functions of  residue position away from the termini
were computed based on the secondary  structure assignment 
recorded in the PDB file (we have discarded structures for which such
an assignment is not present).
In order to better test our prediction we divided our final set of
about 600 proteins into two groups with their relative contact order 
(CO) \cite{Plaxco} greater than and smaller than 0.15. The aim of 
doing this division is to classify
proteins into slow and fast folding, respectively, at least 
in an approximate way
\cite{Plaxco}.  For the former group of proteins the effect we expect should be
less pronounced and indeed we do not see any clear preference from either 
termini (data not shown). On the
contrary the selection effect is  clearly visible in the second group
of proteins characterized by small  CO (Fig. 3a). For this group of proteins 
the helical propensity has a peak in
the vicinity of both termini but the peak closest to the C-terminus is
significantly higher than the one associated with the N-terminus. 
The $\beta$ strand propensity, instead, follows an inverse tendency.
Therefore, the signature of the preferential formation of helices at 
the C-terminal end of a protein, which results
as a consequence of a non-equilibrium 
effect in nascent polypeptides, has been fixed by evolution. 
These results are also compatible with the observation that 
in many proteins translation and folding occur simultaneously
during synthesis \cite{Baldwin}. 
To further support the prediction that non-equilibrium effects are key 
factors leading to the helical preference at the C-terminus we have found 
that the concentrations of the various types of amino acids are similar at 
both the N- and C-termini 20-residue fragments, within the error bars 
provided by the statistics (Fig. 3b).
Fig. 3a also shows that the helical propensity as a
function of the distance from the C terminus displays 
oscillations, which may be either simply due to stochasticity
or to the typical length of a helical segment in the folded
state of proteins.

In conclusion, we suggest that proteins might have taken advantage 
of the dynamical principle for preferential selection of helices, 
described here, to select a polypeptide growth mechanism
that can bias the backbone configuration of a nascent protein into a
state rich in $\alpha$ helices. Our conjecture is supported by
the  existence of a higher propensity of forming helices
near the C-terminus of relatively fast folding proteins.
Furthermore this scenario is consistent with at least two lessons
we know.  First, the reduction of regions rich in $\beta$ sheets
would render protein misfolding and aggregation into amyloid like
structures less likely \cite{dobson}. Second, this particular folding
mechanism would minimize the role of non local contacts, thus
rendering folding faster. It seems appealing that the cell
translation machinery has evolved in such a way as to find an optimal
compromise between the minimisation of errors and the maximisation of
speed in the translation from RNA to nascent proteins. 

We have learnt that A. Laio and C. Micheletti have independently observed the
higher propensity of the C-terminus to form helices. 
This work was supported  by COFIN MURST 2003, EPSRC and the 
Natural Science Council of Vietnam (grant no. 
410704) . We are greatful to the referees for their
very helpful comments.

\begin{figure}
\includegraphics[width=8.cm]{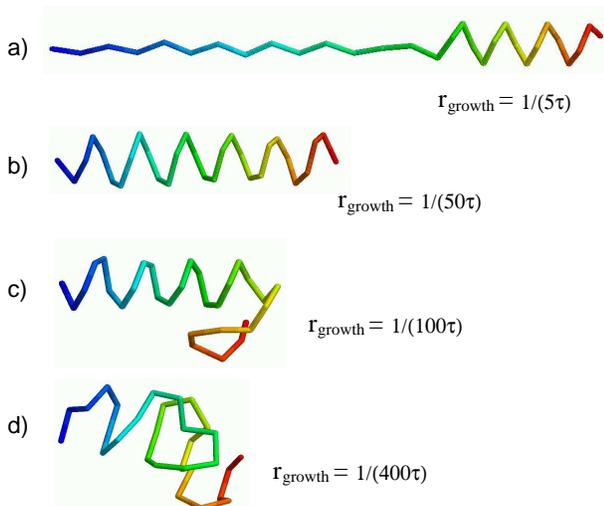}
\narrowtext{
\caption{Conformations of a growing chain for various 
$r_{growth}$ as indicated ($\tau$
is defined in the text).
The polypeptide chains shown are all of 30 residues.
The chain is grown from the left end and the growing end is kept fixed.
The growth rates decrease from top to bottom. 
In a rather wide range of intermediate growth rates (b and c)
an ordered helix is formed
at the growing end and another one, more distorted and equilibrated, 
appears at the other end.
$\tau_{relax}$, at the 
growth temperature, is estimated to be $\sim 80$ $\tau$.}}
\end{figure}

\begin{figure}
\centerline{\includegraphics[width=7.2cm]{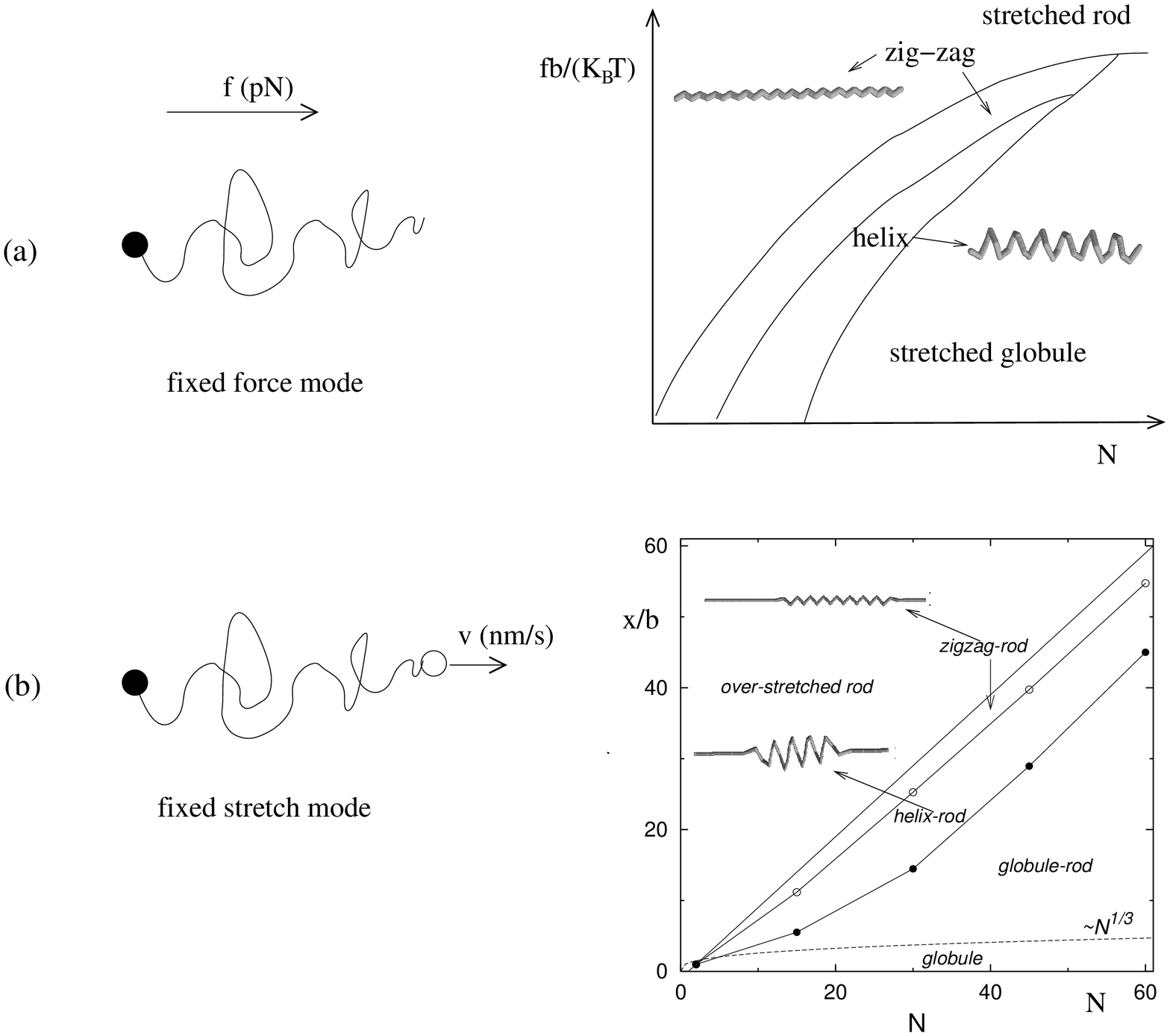}}
\narrowtext{
\caption{Left, schematic representation of two single molecule stretching 
experiments that we discussed:
(a) fixed force mode, (b) fixed stretch mode. Right, sketch of the
 phase diagram at $T=0$. 
The minima were found with molecular dynamics and Monte-Carlo simulations
at low $T$ and varying force (in (a)) and end-to end distance (in 
(b)). In (b), the velocity is small so that the string 
is always in quasi-equilibrium. {Most of the stable states in Fig. 2b
are characterised by a coexistence of two states found in Fig. 2a.}
$N$ refers to the length of the polymer.}}
\end{figure}

\begin{figure}
\centerline{\includegraphics[width=6.3cm]{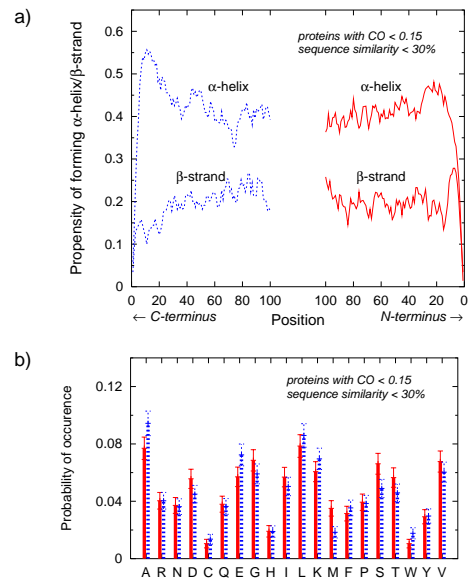}}
\narrowtext{
\caption{(a) Propensity of forming helices and strands as function of residue
position away from C-terminus (blue) and N-terminus (red). The data
are computed based on a PDB set of 291 non-disordered protein structures
with CO smaller than 0.15 and with the secondary
structures assignment given in the PDB files. To determine a
propensity at a given position we only took proteins with length greater
than double the position number, in order to avoid interference from the
opposite end.
(b) The amino acid concentrations in 20-residue fragments associated
with the N- (red) and C-terminus (blue) computed from the same set of
proteins used in the top figure (amino acids are labeled with the one letter
code).}}
\end{figure}

\end{multicols}


\begin{thebibliography}{99}
\bibitem{note} We should compare the growth rate with the rate of 
formation of non-helical structures (among which $\beta$ structures
are the fastest to fold), as the dynamic effect already favors 
helix formation (which occurs in solution in microseconds).
\end{thebibliography}
\end{document}